\documentclass[a4paper,11pt]{article}
\usepackage{jcappub} 
\usepackage{lineno}

\usepackage{aas_macros}

\newcommand{\rsun}{R_\odot}
\newcommand{\msun}{M_\odot}

\newcommand{\gevcc}{{\rm GeV\,cm^{-3}}}
\newcommand{\cc}{{\rm cm^{-3}}}
\newcommand{\kms}{{\rm km\,s^{-1}}}
\newcommand{\rhodm}{\rho_{\rm dm}}
\newcommand{\rhob}{\rho_{\rm b}}

\graphicspath{{./}{./figures/}}

\arxivnumber{2505.17828} 

\title{\boldmath Dark Matter Density Profile Around a Newborn First Star}

\author[a,b,1]{S. Hirano\note{Corresponding author.}}
\author[c,d]{and N. Yoshida}
\affiliation[a]{Department of Applied Physics, Faculty of Engineering, Kanagawa University,\\
3-27-1 Rokkakubashi, Yokohama, Kanagawa 221-0802, Japan}
\affiliation[b]{Department of Astronomy, School of Science, The University of Tokyo,\\
7-3-1 Hongo, Bunkyo, Tokyo 113-0033, Japan}
\affiliation[c]{Department of Physics, School of Science, The University of Tokyo,\\
7-3-1 Hongo, Bunkyo, Tokyo 113-0033, Japan}
\affiliation[d]{Kavli Institute for the Physics and Mathematics of the Universe (WPI), UT Institutes for Advanced Study, The University of Tokyo,\\
5-1-5 Kashiwanoha, Kashiwa, Chiba 277-8583, Japan}

\emailAdd{shingo-hirano@kanagawa-u.ac.jp}

\abstract{
Ambient dark matter (DM) around binary black holes can imprint characteristic signatures on gravitational waves emitted from their merger.
The exact signature depends sensitively on the DM density profile around the black holes.
We run very high resolution cosmological hydrodynamics simulations of first star formation that follow the collapse of a $3\times10^{5}\,\msun$ mini-halo from $z=49$ to $z\simeq22$.
Our flagship model achieves a DM particle mass of $3.7\times10^{-4}\,\msun$ and resolves the inner-most structure down to $0.02$\,pc.
We show that the halo experiences a two-stage gravitational collapse, where a rotating, constant-density core with $r\lesssim3$\,pc is formed first, surrounded by an extended outskirts.
Baryonic infall toward the center continues to raise the local Keplerian velocity and promotes adiabatic contraction of DM.
The resulting density profile has an approximately power-law shape of $\rhodm \propto r^{-0.6}$ inside $\sim 1$\,pc. 
We find that a piecewise power-law fit reproduces the simulation result to better than 10\%, and also find numerical convergence down to $\simeq\!0.01$\,pc.
The DM density profile is typical for ordinary Pop~III halos, but our additional simulations reveal that inner slope varies significantly with halo-to-halo scatter, and the effect of Lyman-Werner irradiation and of supersonic baryon-DM streaming velocities, implying a wide distribution of slopes rather than a single universal curve.
The large variation should be considered when calculating the predicted DM-induced dephasing of gravitational waves by up to an order of magnitude relative to the classical analytic model of the DM spike.
}


\begin{document}
\maketitle
\flushbottom

\section{Introduction} \label{sec:int}

The nature of dark matter (DM) remains an important open question in astrophysics, cosmology, and particle physics.
Although its existence is inferred across virtually all observable length scales, from dwarf galaxies to the cosmic microwave background, the nature of DM is still elusive.
The spatial distribution of DM is critical in many astrophysical processes, including the long-debated formation of core/cusp at galactic scales \cite{Moore1994, deBlok2010, DelPopolo2021}.
Also accurate account of the DM density distribution around astronomical objects is essential for direct and indirect searches of DM.

Gravitational wave (GW) astronomy offers a new avenue to probe fundamental physics \cite{Miller2025review}.
Dynamical friction exerted by ambient DM on a black‑hole (BH) binary perturbs its orbital evolution and imprints a characteristic phase shift on the emitted GW signal \cite{Eda2013}.
If the central BH is surrounded by concentrated DM, dubbed a {\it spike} \cite{GondoloSilk1999}, {\it dress} \cite{Kavanagh2020}, or {\it mound} \cite{Bertone2024, Bertone2025DMmounds}, the accumulated phase shift can be significant.
State‑of‑the‑art waveform calculations suggest that next‑generation facilities are expected to have enough sensitivities to detect the effects \cite{Cole2023PRD,Cole2023NatAs,AmaroSeoane2023LISA}; e.g., third-generation ground-based (ET/CE) and space-based (LISA-like) detectors.
Future GW observatories can thus serve as indirect DM detectors, but the feasibility heavily relies on the model of the DM density profile.

Space‑based missions like LISA will detect extreme and intermediate mass‑ratio inspirals whose central objects have masses $\simeq10^{3}$--$10^{6}\,\msun$ \cite{Colpi2024LISA}.
Observational evidence for intermediate‑mass black holes (IMBHs) has accumulated in nearby dwarf galaxies and globular clusters \cite{Greene2020review}.
At very large masses, the discovery of $\gtrsim10^{9}\,\msun$ quasars at $z\gtrsim7$ continues challenging standard growth scenarios \cite{Inayoshi2020review, Wang2021}.
Recent observations by JWST identified a large number of compact, infrared-bright sources (Little Red Dots; LRDs) that may harbor accreting IMBHs at the center of dark matter halos\cite{Inayoshi2025}.

The first generation of stars (Population III stars; hereafter Pop~III stars) are thought to be formed at the centers of DM minihalos with masses $\sim\!10^{5}$--$10^{6}\,\msun$ at redshifts $z \sim 30$--$15$ \cite{Yoshida2003, KlessenGlover2023}.
Massive Pop~III stars collapse directly into BHs, which can become, under some suitable conditions, as massive as $10^{3}$--$10^{5}\,\msun$.
Because both the baryons and the nascent BH reside deep in the potential well, a dense DM cusp is expected to develop through adiabatic contraction \cite{Blumenthal1986, Gnedin2004}.
Such DM overdensities may substantially affect the GW signal associated with the merger of Pop~III remnant BHs.

Often a simple fitting function is adopted to describe the DM density spike, such as broken power laws based on an initial Navarro-Frenk-White (NFW) profile \cite{NFW1996}.
High‑resolution cosmological simulations including baryonic processes have revealed substantial departures from the self‑similar form \cite{Governato2010}.
There is also a technical issue that several previous studies repeatedly split DM particles during collapse to achieve higher spatial resolution at the risk of artificially changing the inner density slope \cite{Regan2015,Regan2018}.
This motivates us to study the DM profile at the onset of Pop~III star formation.

We present the result of very high resolution cosmological hydrodynamical simulations that resolve the DM distribution inside a minihalo without performing particle splitting from linear density perturbations $z=49$ to gravitational collapse.
Our flagship run (labeled L3) attains a DM particle mass of $m_{\rm dm,min}=3.66\times10^{-4}\,\msun$, and thus improves mass resolution by a factor $\approx 130$ relative to the previous works \cite{Suwa2008,Umemura2012}.
We find that initially a slowly rotating core is formed.
As the baryonic cloud contracts and the local Keplerian velocity increases, the DM core undergoes further contraction, developing a steeper inner slope.
We derive a semi‑analytic formula for the DM density distribution at the Pop~III star formation by fitting the simulation result and by extrapolating to protostellar sizes.
The result provides a realistic input for GW waveform models with the contribution of concentrated DM.

\section{Methods} \label{sec:met}

We generate a hierarchical zoom-in cosmological initial conditions using {\tt MUSIC} \cite{HahnAbel2011} for a volume of $L_{\rm box}=100$\,comoving kiloparsec (ckpc) per side at $z_{\rm ini} = 499$.
We adopt $\Lambda$-Cold Dark Matter ($\Lambda$CDM) cosmology with the cosmological parameters \cite{PLANCK2018}: $\Omega_{\rm m}=0.31$, $\Omega_{\rm b}=0.048$, $\Omega_\Lambda=0.69$, $H_0=68\,\kms\,{\rm Mpc}^{-1}$, and $n_{\rm s}=0.96$.
To accelerate the formation of nonlinear objects (halos) in a small computational domain, we artificially increased the root-mean-square matter fluctuation averaged over a sphere of radius $8\,h^{-1}\,$Mpc from $\sigma_8=0.83$ to $\sigma_8=2$.
This model is dubbed L0, in which the finest particle masses are $m_{\rm dm,min}=0.805\,\msun$ for DM and $m_{\rm b,min}=0.145\,\msun$ for baryon (gas) component.

We perform a set of cosmological simulations with the parallel $N$-body/smoothed particle Hydrodynamics (SPH) code {\tt GADGET-3} \cite{Springel2005}, adapted for metal-free star formation with including detailed non-equilibrium chemistry of 14 species (e$^-$, H, H$^+$, H$^-$, He, He$^+$, He$^{++}$, H$_2$, H$_2^+$, D, D$^+$, HD, HD$^+$, HD$^-$) \cite{Hirano2023FSC1}.
Our simulations use the TreePM gravity of the GADGET family, in which both DM and baryons are treated as discrete particles responding to the same total gravitational potential.
Forces are split into a long-range PM part (FFT on a mesh) and a short-range tree part that is applied uniformly across the domain.
In the high-density, strongly clustered center, the tree walk opens down to leaf nodes with a built-in near-node exclusion zone, so that most near-field interactions are effectively direct particle-particle (DM-DM and DM-baryon) calculations. 
Hence, the short-range interaction in the dense core is not compromised by the multipole approximation; the minimum trustworthy radius is instead set by the adopted (spline) softening length. 
We therefore restrict our discussion and interpretation only down to the gravitational softening length of $\sim\!~0.02$\,pc.

We adopt a two percent of the initial inter-particle distance as the gravitational softening length as $\varepsilon = L_{\rm box} / N_{\rm eff}^{1/3} / 50$.
We run model L0 until the baryon density reaches $\rho_{\rm b}=10^{6}\,\gevcc$, when the primordial gas cloud forms and collapses.
At this stage, the DM minihalo hosting the gas cloud has a virial radius of approximately $R_{\rm virial}\simeq100\,$pc and a mass of $M_{\rm virial}\simeq3\times10^{5}\,\msun$ at redshift $z_{\rm col}\simeq22$, representing the typical properties of a minihalo for the Pop III star formation \cite{Yoshida2003,IshiyamaHirano2025}.

\begin{table}[t]
\centering
\begin{tabular}{c|llll}
\hline
Model & $m_{\rm dm,min}$ & $m_{\rm gas,min}$ & $\varepsilon_{\rm min}$ & $z_{\rm col}$ \\
& ($\msun$) & ($\msun$) & (pc) & \\
\hline
L3   & $3.66 \times 10^{-4}$ & $6.60 \times 10^{-5}$ & $0.022$ & 20.7 \\
L2   & $4.76 \times 10^{-3}$ & $8.58 \times 10^{-4}$ & $0.052$ & 20.8 \\
L1   & $0.0619$              & $0.0112$              & $0.12$  & 21.2 \\
L0   & $0.805$               & $0.145$               & $0.29$  & 22.0 \\
\hline
\end{tabular}
\caption{
Column 1: model name;
Column 2: mass of the finest DM ($N$-body) particle;
Column 3: mass of the finest baryon/gas (SPH) particle;
Column 4: gravitational softening length of the finest DM particle; and
Column 5: redshift when the gas cloud collapses ($\rho_{\rm b,max}=10^{6}\,\gevcc$).
}
\label{tab:models}
\end{table}

Next, we mark the DM and gas particles within the target minihalo and output their positions and other physical quantities at $z=49$.
To enhance the numerical resolution, we perform particle splitting on DM and gas particles within a region that sufficiently encompasses their spatial distribution, roughly twice the final virial radius.
We adopt the particle splitting technique described in \cite{KitsionasWhitworth2002}, which arranges the $13$ child particles on a hexagonal close-packed array.
Using the same L0 snapshot at $z=49$ as a common progenitor, we build three independent higher-resolution initial-condition sets by splitting each DM and gas particle into $13^n$ child particles ($n=1,2,3$).
We then evolve these three sets separately as models L1–L3 (table~\ref{tab:models}) until the target gas cloud collapses ($\rhob=10^{6}\,\gevcc$).
The highest-resolution model (L3) achieves a mass resolution of DM component $m_{\rm dm,min} = 3.66\times10^{-4}\,\msun$ with gravitational softening of $\varepsilon_{\rm min}=0.022\,$pc.

\begin{figure}[t]
\centering
\includegraphics[width=0.8\textwidth]{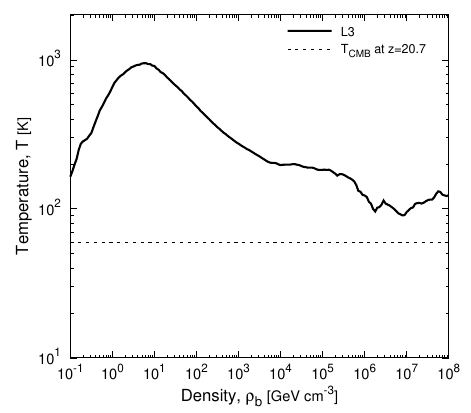}
\caption{
Averaged gas temperature of the collapsing cloud as a function of the gas particle number density for Model L3.
The horizontal dotted line shows the CMB temperature floor $T_{\rm CMB}(z)=2.73(1+z)\,$K at $z=20.7$.
Note the effect of HD cooling at $\rho_{\rm b} > 10^5$.
The temperature does not increase gradually unlike in the usual case with H$_2$ cooling only. 
}
\label{fig:Dens-Temp_L3}
\end{figure}

We continue the simulations until the gas density reaches $\rho_{\rm b,max}\simeq10^{9}\,\gevcc$, which corresponds to the numerical resolution limit where the Jeans length falls below 10 times the SPH particle’s smoothing length.
The Jeans length is determined by the local density and temperature of the gas particles (figure~\ref{fig:Dens-Temp_L3}).
We then analyze the final and intermediate outputs to study the evolution of the DM density profile during the cloud collapse within the minihalo.

The radial profiles shown in section~\ref{sec:res} are spherically averaged, and regions with fewer than 100 cumulative particles are excluded.
This effectively sets our conservative spatial resolution.
When calculating DM density profiles, we define the center at the location where the {\it total} (DM and baryon) density peaks.  
We adopt the total-density center to probe the DM distribution near the baryonic density peak, where Pop~III stars and remnant black holes are left.\footnote{Density profiles defined around the {\it center} may, in practice, differ from those centered strictly at the peak of the DM density. Our direct comparison confirmed, however, that the two centers almost coincide, and the resulting discrepancies are negligible.}

\section{Numerical simulations} \label{sec:res}

\begin{figure*}[t]
\centering
\includegraphics[width=1.0\textwidth]{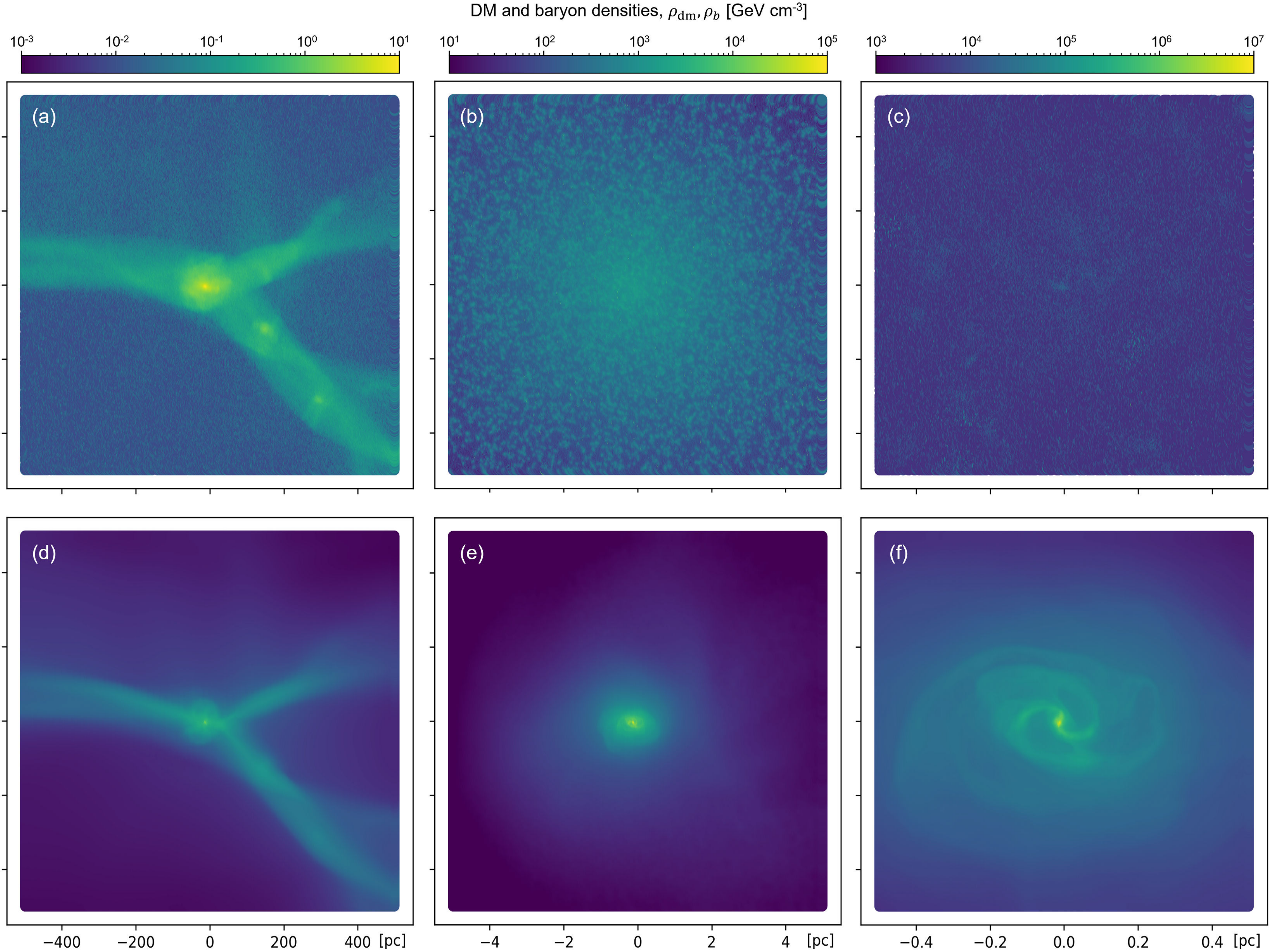}
\caption{
Density distributions of DM ($\rho_{\rm dm}$; top panels) and baryonic ($\rho_{\rm b}$; bottom panels) components projected on a cross-section through the density center.
The left, center, and right panels show plots over $1000$, $10$, and $1$\,pc per side, respectively.
}
\label{fig:map2d_L3}
\end{figure*}

We perform cosmological simulations until the first gas cloud collapses in the simulation box.
Figure~\ref{fig:map2d_L3} shows the projected density distributions of the DM and baryon components ($\rhodm$ and $\rhob$) at the end of the simulation ($z=20.7$) of the highest-resolution model (L3).
Clearly, on scales of $\sim 1000$\,pc (figures~\ref{fig:map2d_L3}a and d), large-scale structures are evident, and the DM density exceeds the baryon density ($\rhodm > \rhob$).
The gas is dragged or trapped by the gravitational potential of DM, and forms essentially the same, coherent structure.
At scales of $10$\,pc within the nonlinear halo (figures~\ref{fig:map2d_L3}b and e), the DM density appears roughly constant.
In contrast, the baryon density increases sharply toward the center, generating a ``self-gravitating'' density distribution ($\rhodm < \rhob$).
DM shows a flat density distribution at the inner region of $1$\,pc (figures~\ref{fig:map2d_L3}c and f), whereas the gas further contracts by radiative cooling.
We have found that the prominent spiral arms and arm-like structures are formed when the gas temperature drops owing to HD cooling (see figure~\ref{fig:Dens-Temp_L3}). Formation of and cooling by HD molecules are promoted by the slow collapse of the rotating gas cloud (see appendix~C of \cite{Hirano2014} and figure~9 of \cite{Hirano2015}).

We study the radial profiles of a variety of physical quantities using the simulation outputs.
In this section, we present the results of the highest-resolution model (L3), which serves as the main result of the present paper.
We assess the numerical convergence of the simulation results in appendix~\ref{app:convergence}.

\begin{figure}[t]
\centering
\includegraphics[width=1.0\textwidth]{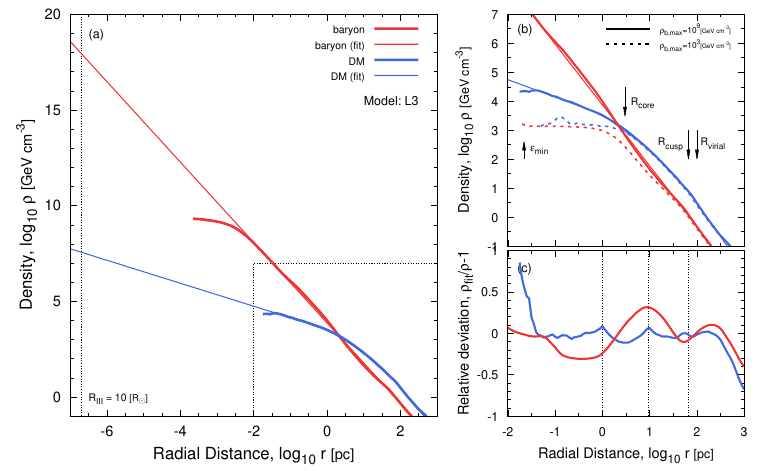}
\caption{
Radial density profiles of the DM and baryonic components for Model L3.
Panel (a): the thick lines are simulation results when the maximum baryon density is $\rho_{\rm b,max}=10^9\,\gevcc$ and the thin lines are fitting functions for the two components (eqs.~\ref{eq:rhodm_fit} and \ref{eq:rhob_fit}). 
The vertical dotted line is $R_{\rm III}=10\,\rsun$, the radius of the typical Pop~III star with mass $M_{\rm III} = 100\,\msun$ \citep{HosokawaOmukai2009}.
Panel (b): enlarged view of the range indicated by the dotted rectangle in panel (a).
The dashed lines are simulation results when baryon density exceeds DM density, $\rho_{\rm b,max}=10^3\,\gevcc$.
Arrows indicate characteristic radii of the DM density profile: virial radius of the DM halo ($R_{\rm virial}$), scale radius where DM density profile steepens ($R_{\rm cusp}$), core radius at which the DM density flattens before adiabatic contraction due to gas condensation ($R_{\rm core}$), and gravitational softening length of the finest DM particle ($\varepsilon_{\rm min}$).
Panel (c): relative deviation of our fitting functions from the simulation results.
The vertical dotted lines are the radii at which the fitting function of the DM component switches, $r=1.0, 9.4, 66$\,pc, respectively.
}
\label{fig:Radi-Dens_L3-multi}
\end{figure}

\subsection{Radial density profile}

The key results are the radial density profiles of DM and baryonic components for model L3.
Figure~\ref{fig:Radi-Dens_L3-multi}(a) plots the final radial profiles (thick lines) obtained from the simulation over a radial range spanning $12$ orders of magnitude, from the halo scale (kpc) down to the surface of the first star ($R_{\rm III}=10\,\rsun$).
At large radii, the DM density exceeds the baryon density by a constant factor of approximately $(\Omega_{\rm m}-\Omega_{\rm b})/\Omega_{\rm b}\simeq5.5$ (see also Figure~\ref{fig:map2d_L3}).
The gas density follows nearly a constant power-law, close to the isothermal profile, but the slope of the DM density profile decreases with decreasing radius; baryons are dominant within $r\simeq3\,$pc.

Figure~\ref{fig:Radi-Dens_L3-multi}(b) is an enlarged view of panel (a).
For DM, the profile becomes progressively shallower toward the center, with the power-law index varying multiple times.

\begin{figure}[t]
\centering
\includegraphics[width=0.8\textwidth]{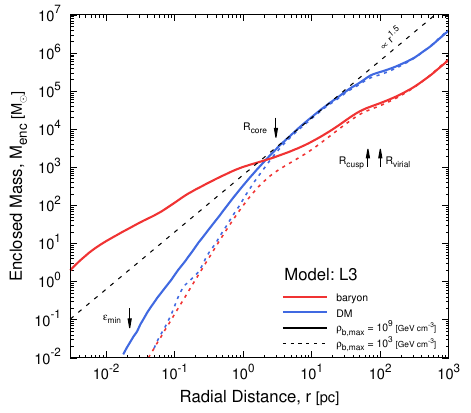}
\caption{
Radial profiles of the enclosed masses of baryon and DM components (red and blue lines) for Model L3 when $\rho_{\rm b,max}=10^3$ and $10^9\,\gevcc$ (dashed and solid line styles).
The long-dashed line shows a correlation as $M_{\rm enc} \propto r^{1.5}$.
}
\label{fig:Radi-Menc_L3}
\end{figure}

\subsection{DM cusp at $R_{\rm core} < r < R_{\rm cusp}$}

The DM density profile gets shallower at $R_{\rm cusp}=66$\,pc, and a similar transition is found in the baryon density profile as well.
This is typical for a gas within an NFW halo.
The cumulative DM mass shows that within the cusp radius $R_{\rm cusp}$, the DM mass scales approximately as $M_{\rm enc,dm}(r) \propto r^{1.5}$ (figure~\ref{fig:Radi-Menc_L3}).

\subsection{DM core at $r < R_{\rm core}$}

Further inward, the power-law slope changes again at $R_{\rm core} \sim 3$\,pc.
The dashed lines in figure~\ref{fig:Radi-Dens_L3-multi}(b) represent the DM and baryon density profiles at the time when $\rho_{\rm b,max}=10^3\,\gevcc$.
At this time, both DM and the gas density profiles exhibit a constant density ``core'' with $\rhodm \simeq \rhob$.
Subsequently, the gas density far exceeds the DM, and continues condensing by $H_2$ and HD cooling, and the DM density profile steepens again to have a cusp at $r < R_{\rm core}$.

\begin{figure}[t]
\centering
\includegraphics[width=0.8\textwidth]{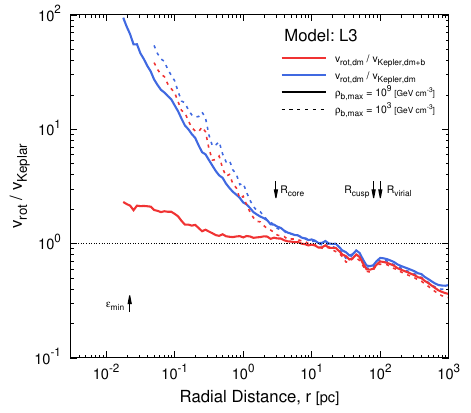}
\caption{
Radial profiles of the rotation velocity with respect to the Keplerian velocity for Model L3 when $\rhob=10^3$ and $10^9\,\gevcc$.
The red lines show the ratio to the Keplerian velocity obtained from the total mass, $v_{\rm Kepler,dm+b} = \sqrt{G(M_{\rm enc,dm}(r)+M_{\rm enc,b}(r))/r}$, whereas the blue lines show the ratio to the Keplerian velocity obtained from the DM mass only, $v_{\rm Kepler,dm} = \sqrt{GM_{\rm enc,dm}(r)/r}$.}
\label{fig:Radi-VrotVkep_L3}
\end{figure}

Before the onset of gravitational run-away collapse of the primordial gas cloud, a flat DM ``core'' with $\sim\!1$\,pc develops.
As shall be discussed later in section~\ref{sec:res:kinematic}, the DM core consists of low velocity and low angular momentum materials which, however, cannot contract infinitesimally without any dissipation process for DM.
Figure~\ref{fig:Radi-VrotVkep_L3} plots the rotational-to-Keplerian velocity ratios calculated from the total mass, $v_{\rm rot}/v_{\rm Kepler,dm+b}$ (red), and from the DM mass alone, $v_{\rm rot}/v_{\rm Kepler,dm}$ (blue).
At the early stage ($\rho_{\rm b,max}=10^3\,\gevcc$; dashed) both ratios exceed unity for $r < R_{\rm core}$.
The near Keplerian rotation prevents further DM infall and sets the cored profile.\footnote{This mechanism is different in origin from the core formation induced by baryonic feedback discussed in the context of galaxy formation.}
As baryons continue to contract (by dissipation), their cumulative mass inside $R_{\rm core}$ grows ($\rho_{\rm b,max}=10^9\,\gevcc$; solid).
Consequently, $v_{\rm rot,dm}/v_{\rm Kepler,dm+b}$ falls to $\approx 1$, the DM particles experience adiabatic contraction, and the inner slope steepens once more \cite{Blumenthal1986}.
Since $v_{\rm rot,dm}/v_{\rm Kepler,dm}$ remains $> 1$, it is evident that the DM collapse is induced solely by the baryon's gravitational potential.

In the late, high-density phase, the DM inside the previously formed cored profile undergoes a baryon-induced re-contraction: as the dissipative baryon collapse deepens the central potential, the DM responds adiabatically.
Because DM lacks dissipation, this re-contraction is subordinate to the baryon collapse rather than self-driven.
Consequently, the DM density center follows the baryon density center, and within our spatial resolution, the two peaks coincide during this phase (see figure~\ref{fig:Radi-Dens_L3-multi}).
This confirms in our simulations that a DM core established at earlier times can steepen again once the baryon collapse proceeds.

\begin{figure}[t]
\centering
\includegraphics[width=1.0\textwidth]{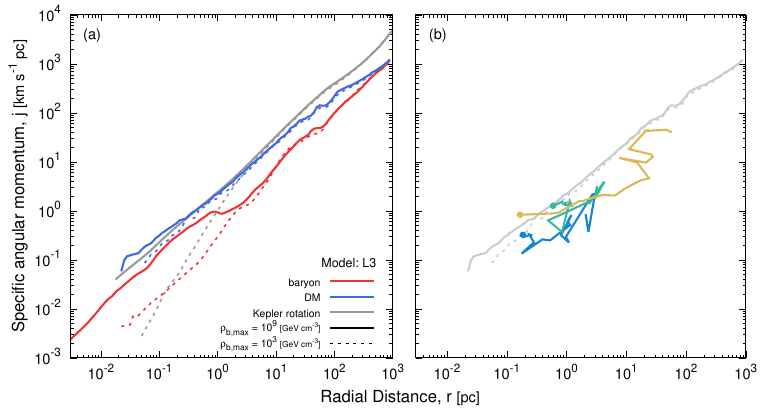}
\caption{
Radial profiles of the specific angular momentum
Panel (a): the specific angular momentum of baryon and DM components for Model L3 when $\rhob=10^3$ and $10^9\,\gevcc$.
The gray lines show the specific angular momentum in the case of Keplerian rotation as $v_{\rm Kepler,dm+b} = \sqrt{G(M_{\rm enc,dm}(r)+M_{\rm enc,b}(r))/r}$.
Panel (b): time evolution of the specific angular momentum of the selected three DM ($N$-body) particles calculated around the maximum density coordinates at each time.
The filled circles represent the final states for three particles at the end of the simulation.
The gray solid and thin lines show the radial profile of the DM component (the blue lines in panel a).
}
\label{fig:Radi-Angmom_L3}
\end{figure}

\subsection{Kinematic support and angular-momentum evolution} \label{sec:res:kinematic}

Figure~\ref{fig:Radi-Angmom_L3}(a) shows the radial profile of the specific angular momentum, $j(r)$.
The colored curves show mass-weighted means for DM (blue) and baryons (red), whereas the gray curve indicates the Keplerian value $j_{\rm Kepler}(r) = \sqrt{G M_{\rm enc}(r) r}$.

Inside the core radius ($R_{\rm core}$), the DM curve closely traces $j_{\rm Kepler}$, indicating that rotational support prevents further collapse.
On the halo, kilo-parsec scale, baryons share the same $j$ as DM, but toward smaller radii the profile gets steeper and the specific angular momentum is smaller by $\simeq 0.5$\,dex because kinetic energy is partially dissipated by radiative cooling.
A second break appears at $\sim 1$\,pc, which corresponds to the gas cloud size that contracts rapidly by H$_2$ and HD cooling (see figure~\ref{fig:Dens-Temp_L3}).

It is important to examine whether individual DM particles lose angular momentum during infall.
Figure~\ref{fig:Radi-Angmom_L3}(b) traces the histories of three representative DM particles.
We plot the time evolution of the specific angular momentum calculated around the maximum density coordinates at each time.
We find that the particles that ultimately reach the inner $0.1$\,pc region start with relatively low $j$, and indeed evolve while conserving angular momentum.
In contrast, particles that originally had higher $j$ remain at larger radii.
We argue that, for DM, the stratification in the inner angular momentum profile results from the initial values at the time of gravitational collapse. 
This behavior is reminiscent of the ``pre-selection'' of low-$j$ particles reported in galaxy-scale simulations.
\cite{Zavala2016} showed that the inner DM particles later residing within 10\% of the virial radius of a $z \approx 0$ halo already possessed lower specific angular momentum near turnaround and subsequently evolved almost adiabatically, while high-$j$ particles remained at larger radii.
Our result extends their picture down to $\sim10^{5}\,\msun$ halos at $z\gtrsim20$, suggesting that angular-momentum stratification is a scale-free consequence of tidal torque, rather than of late-time dynamical friction or baryonic feedback.

\subsection{Fitting functions}

We find that the DM density profile exhibits multiple power-law segments, whereas the baryon density profile maintains an almost constant power-law.
We fit these radial density profiles as follows:
\begin{align}
  \label{eq:rhodm_fit}
  \rho_{\rm dm,fit} &= \rho_{\rm i} \left( \frac{r}{{\rm pc}} \right)^{-\gamma_{\rm i}} \,\gevcc \hspace{2mm}(i=1,2,3,4) \,, \nonumber\\[1ex]
  &\quad\text{with}\quad
  \begin{cases}
    \{ \rho_{i} \}   = \{ 1.2\times10^5, 1.2\times10^4, 3.5\times10^3, 3.5\times10^3 \} \,\gevcc \,, \\
    \{ \gamma_{i} \} = \{ 2.3, 1.75, 1.2, 0.6 \} \, ,
  \end{cases}\\
  \label{eq:rhob_fit}
  \rho_{\rm b,fit} &= \rho_{0} \left( \frac{r}{{\rm pc}} \right)^{-\gamma_{0}} \,\gevcc \,, \nonumber\\[1ex]
  &\quad\text{with}\quad
  \begin{cases}
    \rho_0   = 7.5\times10^3 \,\gevcc \,, \\
    \gamma_0 = 2.1 \, .
  \end{cases}
\end{align}
We overlay the best-fit models $\rho_{\rm dm,fit}$ and $\rho_{\rm b,fit}$ (thin lines) for comparison in figure~\ref{fig:Radi-Dens_L3-multi}.
Panel (c) in figure~\ref{fig:Radi-Dens_L3-multi} shows the relative deviation between the fitting functions and the simulation results.
The relative deviation in the DM density is confined within $\pm10\%$.
Although the relative deviation for the baryon density is somewhat larger (within $\pm30\%$), this is likely due to the presence of substructure (see right bottom panel of figure~\ref{fig:map2d_L3}).

\begin{figure}[t!]
\centering
\includegraphics[width=0.8\textwidth]{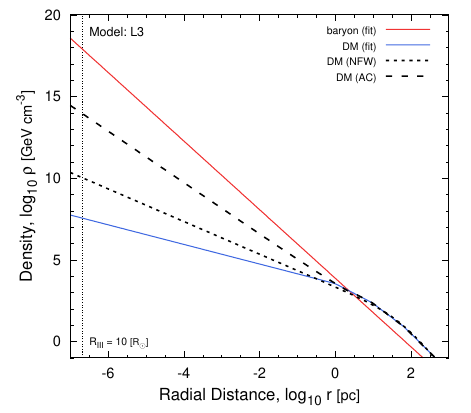}
\caption{
Comparison of the DM density profiles obtained by the simulation (L3; solid) and theoretical estimations: the dotted line is the Navarro-Frenk-White (NFW) profile \cite{NFW1996} and the dashed line is the profile obtained assuming the adiabatic contraction (AC) \cite{Blumenthal1986}.
}
\label{fig:Radi-Dens_L3-theory}
\end{figure}

\section{Discussion}\label{sec:dis}

\subsection{Comparison with theoretical models}\label{sec:disc_models}

We have calculated the DM density profile inside a Pop III star-forming minihalo in very high-resolution cosmological simulations.
Comparing this simulation-based profile with classical analytical models is essential for devising accurate models of dark matter spikes, decay and annihilation effects, and GW signature.
Figure~\ref{fig:Radi-Dens_L3-theory} compares the L3 fitted profile and two standard theoretical models.
We can better understand potential biases in previous GW analyses that assumed idealized halo shapes by examining how the simulation's DM distribution differs from the well-known profiles.

We consider two benchmark DM density models used in GW studies and compare them to our L3 simulation profile:
\begin{enumerate}
    \item Navarro-Frenk-White (NFW) profile:
    Using the same scale radius and normalization as the L3 fit for an apples-to-apples comparison, we find that the NFW profile \cite{NFW1996} is in good agreement with the simulation at larger radii (outside roughly 10\,pc; see figure~\ref{fig:Radi-Dens_L3-theory}).
    In the outer halo, the L3 density profile is close to NFW, indicating that the overall halo concentration is similar.
    However, NFW diverges in the inner region because of its fixed inner cusp slope of $\gamma = 1$.
    This is steeper than the $\gamma \approx 0.6$ inner slope obtained from the L3 simulation fit. Consequently, adopting the NFW model overestimates the DM density in the central halo.
    In other words, as one moves toward the halo center (within $\sim\!1$\,pc and especially at sub-pc scales), the NFW curve rises more sharply than the L3 profile.
    This suggests that a naive application of NFW would predict more DM mass near the protostar than in our simulation.

    \item Adiabatic contraction (AC) model:
    We also compare the simulation result with the adiabatic contraction model  \cite{Blumenthal1986}.
    To this end, we assume that the DM halo initially had the density profile given by the L3 simulation when the central gas density reached $\rhob = 10^3\,\cc$ (an early stage of the protostellar collapse).
    We then apply the AC formalism using the final baryonic density profile from the L3 run, i.e., the gas distribution at the time of Pop~III star formation, to compute how the DM would respond to the enhanced gravitational potential.
    The resulting DM profile develops an extremely steep central cusp.
    We find an effective inner slope of $\gamma \sim 1.6$, much steeper than the NFW and L3 profiles. 
    This indicates that, under the assumption of adiabatic contraction, the infall of baryons would drag DM to form a high-density ``spike'' near the center.
    Indeed, the AC model dramatically overshoots the simulated DM density in the inner halo.
    As shown in figure~\ref{fig:Radi-Dens_L3-theory}, the AC curve rises far above the L3 curve at small radii at the time of the snapshot.
\end{enumerate}
Quantitatively, at the radius corresponding to a newly formed Pop~III star’s surface ($R_{\rm III} \approx 10\,\rsun$), the NFW and AC models predict DM densities roughly $10^2$ and $10^{6.5}$ times higher, respectively, than that given by the L3 simulation fit.
These differences underscore how strongly the choice of DM profile can affect estimates of the inner halo density around the star.

The above comparisons highlight that the L3 simulation profile is considerably shallower in the core than the classical models, which has important implications for GW templates that include DM effects.
It is important to emphasize that the L3 profile represents the DM distribution {\it at the moment of Pop~III star formation}, essentially a snapshot when the protostar has just formed at the halo center.
However, the actual GW emission from any remnant black hole binaries would occur later, after the stellar evolution, collapse, and growth of the remnant BH.
The DM structure could continue to evolve while the central star grows by accreting the surrounding gas. This may last at least for $10^5$ years.
While the L3-derived profile is a more realistic baseline than simplistic NFW or AC assumptions at $t=0$ of star birth, one should be careful with how to use the result for GW waveform predictions.
We can only fully assess the impact on GW signals by tracing the DM profile forward in time, from the moment of star formation to the epoch of black hole inspiral.

\subsection{Variety of the DM density slope}\label{sec:dis_comp}

\begin{figure*}[t!]
\centering
\includegraphics[width=1.0\textwidth]{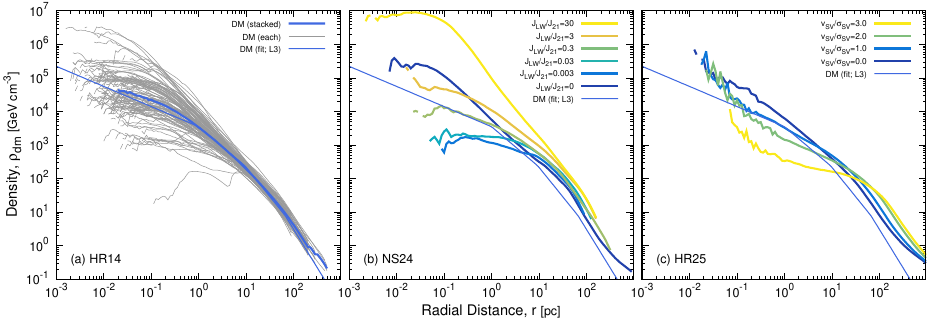}
\caption{
Comparison of DM density profiles among the fitting function of Model~L3 (eq.~\ref{eq:rhodm_fit}) and simulation results in the previous works with considering (a) different halos \cite{Hirano2014} (b) external Lyman-Werner radiation $J_{\rm LW}$ \cite{Nishijima2024}, and (c) initial streaming velocities $v_{\rm sv}$ \cite{Hirano2025FSC2}.
}
\label{fig:Radi-Dens_COMP-multi}
\end{figure*}

We note that the fitting function derived from our L3 model is for just a typical Pop~III host, and there may likely be substantial variation between different halos (Pop~III hosts).
Figure~\ref{fig:Radi-Dens_COMP-multi} shows the diversity of the DM density profiles in Pop~III star-forming halos.
The density slope is nearly identical across all halos from the virial radius down to the cusp scale ($\gtrsim10$\,pc).
However, the profiles diverge markedly at smaller radii from one halo to another.
The three panels highlight three distinct physical mechanisms that generate the dispersion in the inner DM density slope.

\begin{enumerate}
    \item[(a)] {\it Different halos}: We compare the L3 model with 110 distinct minihalos that host Pop~III stars and have virial masses $M_{\rm virial}\simeq10^{5}$--$10^{6}\,\msun$ \cite{Hirano2014}.
    The stacked profile, obtained by taking the logarithmic mean of all DM density profiles, is almost identical to that of the L3 model.
    This close agreement suggests that the fitting function represents a typical Pop~III density structure.
    In contrast, the DM density slopes of individual halos exhibit significant scatter at radii $\lesssim10$\,pc, especially in regions where the gas number density exceeds $100\,\cc$.
    
    \item[(b)] {\it Radiative effect}: Radiation from nearby stars and galaxies alters the chemical composition of star-forming gas clouds, thereby changing the conditions required for the onset of self-gravitating collapse, such as the critical halo mass.
    For instance, when Lyman-Werner (LW) radiation is present, it photodissociates molecular hydrogen, the primary coolant in primordial gas, so the cloud can collapse after its halo grows to a much larger mass.
    Panel~(b) displays the DM density profiles at the moment of gas collapse for the same cosmological halo irradiated by different LW intensities $J_{\rm LW}$ \cite{Nishijima2024}.
    Note that the simulations are run for an identical halo.
    The external radiation field affects the cooling and condensation of the gas, which dramatically reshapes its DM density profile.
    In the so-called direct-collapse regime with $J_{\rm LW}/J_{21}=30$, which is currently a leading scenario for forming early black holes, H$_2$ formation is completely suppressed, the gas remains warm until efficient atomic cooling induces an almost monolithic collapse.
    It is expected that the system produces a supermassive-star progenitor embedded in a very steep DM density cusp.
    
    \item[(c)] {\it Cosmological streaming velocity}: The supersonic relative motion between DM and baryons imprinted at cosmic recombination is another external factor that influences Pop~III star formation \cite{Tseliakhovich2010, Fialkov2014}.
    The streaming velocity (SV) typically delays the onset of gas collapse, promotes star formation in more massive DM halos, and raises the characteristic Jeans mass.
    Panel~(c) presents the DM density profiles of halos formed under four initial streaming velocities ($v_{\rm SV}/\sigma_{\rm SV}=0$, 1, 2, and 3) \citep{Hirano2025FSC2}.
    Halos that form in regions with larger SV collapse later attain greater virial masses, yet display lower central DM densities.
    This trend likely arises because the coherent, megaparsec-scale relative flow tends to smooth out the small-scale structure inside the halo.
\end{enumerate}

Overall, the inner DM slope is determined by a tug-of-war between cooling-driven gas infall and other physical processes, such as radiative heating that delays or displaces the gas.
The primordial minihalos that form ordinary Pop~III stars, supermassive stars, or DCBHs produce different DM density structures as sketched in figures~\ref{fig:Radi-Dens_COMP-multi}.
Clearly, a systematic parameter study is needed to make the above qualitative investigation to theoretical prediction for early black hole formation and the DM distribution.

\section{Conclusion}\label{sec:con}

We have examined the DM density profile of Pop~III star-forming halos by performing cosmological hydrodynamics simulations.
We have followed how the DM distribution responds to the collapse of a primordial gas cloud and rearranges the DM density distribution.

We summarize our findings as follows.
\begin{enumerate}
    \item {\it How does the cloud collapse reshape the DM density structure?}:
    A three-layer DM structure is formed by halo formation and gas condensation.
    On halo scales ($\gtrsim10$\,pc), the DM obeys an NFW-like cusp and outweighs the gas by the cosmological baryon-to-DM ratio.
    However, radiative cooling lets baryons plunge inward, invert the density order, and flatten the DM slope.
    The profile first shallows at the cusp radius ($66$\,pc) and transitions to a genuine constant density core at $\sim\!3$\,pc, where super Keplerian rotation traps DM particles and arrests further inflow; specific angular momentum profiles show DM hugging the local Keplerian value while high-$j$ particles stall at larger radii.
    Continued gas accretion deepens the central potential, cancels the rotational excess, and restarts adiabatic contraction, steepening the DM cusp by protostellar densities.
    Cloud collapse imprints a transient, rotation-supported core whose eventual erosion links the inner halo’s density evolution to the competing influences of collisionless dynamics and rapidly cooling baryonic infall.

    \item {\it Is the resulting profile robust to resolution?}:
    Multi-resolution tests (L0-L3) show that the dark-matter profile stabilizes long before numerical limits are reached: successive 13 times increases in particle number leave the core radius, inner slope, and overall multi-power-law shape essentially unchanged, and even the shallowest segment ($\approx0.01$--$0.1$\,pc) differs by only a few percent between runs.
    A four-segment fit reproduces the L3 profile to $\pm10\%$ and the baryonic profile to $\pm30\%$, the latter scatter arising from small-scale substructure rather than resolution errors.
    Because the central density and slope asymptote to the same values as resolution improves, the flattening and subsequent steepening of the cusp seen in L3 are physical, not numerical, confirming that our key conclusions are robust against variations in particle mass and softening length.

    \item {\it Is the fitting function universal?}:
    The fitting represents a typical Pop III DM profile, but figure~\ref{fig:Radi-Dens_COMP-multi} clearly shows that it is not universal.
    The DM distribution remains almost identical from the virial radius to roughly $10$\,pc, but diverges because of at least three effects: halo-to-halo variations, LW backgrounds, and supersonic streaming velocity.
\end{enumerate}

These findings highlight a complex behavior of DM distribution at the smallest cosmological scales: gravity and angular momentum first form a temporal core and later rebuild a cusp.
Our result provides a renewed link between high‑redshift structure formation and upcoming GW observations with physically motivated, resolution-tested density profiles.
Incorporating these profiles into BH‑binary inspiral models will refine forecasts for DM‑induced dephasing, and future detection or even a null result will give a fresh insight into the DM distribution in the first halos, possibly offering a probe of the nature of DM.

\appendix

\begin{figure}[t!]
\centering
\includegraphics[width=0.8\textwidth]{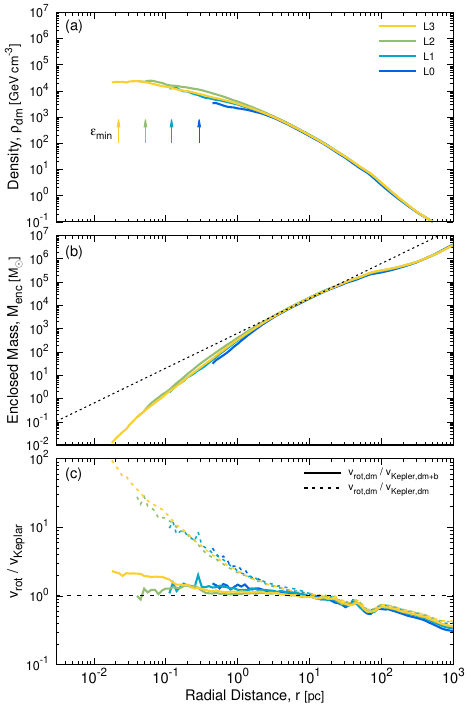}
\caption{
Radial profiles for Models L0, L1, L2, and L3 (see table~\ref{tab:models}).
Panel (a): DM density (as figure~\ref{fig:Radi-Dens_L3-multi}).
Panel (b): enclosed DM mass (as figure~\ref{fig:Radi-Menc_L3}).
Panel (c): rotational-to-Keplerian velocity ratios of the DM component (as figure~\ref{fig:Radi-VrotVkep_L3}).
}
\label{fig:L0toL3}
\end{figure}

\section{Numerical convergence}\label{app:convergence}

We evaluate the numerical convergence of DM profiles by repeating the simulation at four mass resolutions (L0-L3; see Figure~\ref{fig:L0toL3}).
Each step increases the DM particle count by a factor of $13$ (table~\ref{tab:models}).

The density profiles obtained with our highest resolution model (L3) agree with those from lower-resolution ones, except very close to the resolution limit.
By resolving the DM minihalo with progressively higher particle counts, the inferred central density and slope asymptote to stable values.
The change of the DM density slope in model L3 is therefore a physical result, not a numerical artifact.
Indeed, the shallow density structure that develops (of order $\sim 0.01$--$0.1$\,pc in radius) persists with only minor differences when we vary the mass resolution or softening length.

The multi-resolution test demonstrates that model L3 captures the DM structure around the Pop~III star, lending robustness to all key results presented in this paper.

\acknowledgments
The authors thank Jonathan Tan and Jesus Zavala for fruitful discussion.
Numerical computations were performed on Cray XC50 and XD2000 at CfCA in the National Astronomical Observatory of Japan and Yukawa-21 at YITP in Kyoto University.
Numerical analyses were carried out on analysis servers at Center for Computational Astrophysics, National Astronomical Observatory of Japan.
This work made use of {\tt gnuplot}, {\tt numpy} \cite{NumPy2020} , {\tt scipy} \cite{SciPy2020}, and {\tt matplotlib} \cite{Matplotlib007} software packages.
This work was supported by JSPS KAKENHI Grant Numbers JP21H01123 and JP23K20864 (S.H.).


\bibliographystyle{JHEP}
\bibliography{ms.bib}

\end{document}